\documentclass[aps,11pt,prd,superscriptaddress,preprintnumbers,
	amsfont,
	amssymb,
	amsmath,
	notitlepage,
	longbibliography,
	nofootinbib]{revtex4-1}
\usepackage[colorlinks=true, 
	linkcolor=blue, 
	citecolor=blue, 
	urlcolor=blue, 
	linktocpage=true]{hyperref}
\usepackage[english]{babel}
\usepackage{graphicx}

\begin{document}

\title{Effective field equations with stochastic initial conditions}
\date{\today} 

\author{Ian G. Moss}
\email{ian.moss@newcastle.ac.uk}
\affiliation{School of Mathematics, Statistics and Physics, Newcastle University, 
Newcastle Upon Tyne, NE1 7RU, UK}

\begin{abstract}
The evolution of quantum field expectation values from an initial non-equilibrium quantum 
state can be followed numerically
using the Truncated Wigner technique. It can track non-perturbative quantum phenomena,
such as false vacuum decay, in real time with promising results. Nevertheless, it is an approximate method
subject to quantum loop corrections. The extension of Truncated Wigner into a fully
fledged effective field equation approach is discussed here, including corrections
from higher loops and higher momenta.
\end{abstract}

\maketitle

\section{Introduction}

Suppose we would like to follow the time evolution of expectation values for quantum fields.
In conventional quantum field theory, we can introduce an affective action to generate equations 
for field expectation values \cite{Coleman:1973jx}. This works well for finding equilibrium states
and responses to external sources, but it is less well-suited to non-equilibrium states. For
non-equilibrium systems we would want to look at the evolution of the density
matrix, or the related Wigner functional, but this is often impractical in quantum field theory.
By contrast, the Truncated Wigner approach, used widely in condensed matter theory,
offers a practical approach \cite{Blakie}.
Initial conditions are sampled from a statistical distribution and the evolution is followed {\it classically}. 
It seems plausable that the two approaches, Truncated Wigner and the effective action,
should to be combined, and the technical details of how to do this are discussed here.

The Truncated Wigner formalism is mostly useful for numerical studies of a quantum system that
starts out of equilibrium. It is being increasingly adopted
to study highly non-linear quantum phenomena such as false vacuum decay
\cite{Braden:2017add,Braden:2018tky,Braden:2019vsw,Billam:2018pvp,Billam:2021nbc,Hertig:2026oav}. 
An important question in these
kind of studies is how to incorporate quantum loop effects. They have been included 
in some studies by integrating out
a set of coupled fields, but quantum corrections arising from the stochastic field itself are
less well understood. So far, this has mostly been limited to looking at the effects of mass
renormalisation \cite{Braden:2022odm}. Methods exist that go beyond the Truncated Wigner
Approximation \cite{PhysRevA.68.053604,Polkovnikov_2010},
but these do not use loop expansions, and they have mostly been applied in discrete quantum
systems.

Truncated Wigner methods are complementary to the Langevin methods widely used in thermal field
theory (for example, see 
\cite{Gleiser:1993ea,GardinerStochastic2002,GardinerStochastic2003,Boyanovsky2015}). 
The Langevin approach is adapted from the theory of open-systems (see \cite{Breuer:2007juk}), 
where integrating out thermal fluctuations (or noise), from a coupled system, introduces a stochastic 
{\em source} term into the field equations \cite{Feynman1963}. By contrast, Truncated Wigner methods 
seek to represent the initial state by stochastic initial conditions. Nevertheless, the two approaches 
have similarities, and can be combined. This will be discussed in the concluding section.

We use a complex field representation that applies equally 
well to relativistic and non-relativistic systems, and allows a more rigorous treatment 
of initial conditions than can be obtained using real fields. The field satisfies an
effective field equation which includes coupled field and self-field quantum corrections. This equation
can be derived from an effective action in the in-in or Schwinger-Keldysh formalism.
The punchline is this - the self-field one-loop corrections are absent when extending the Truncated Wigner method. 
Corrections to Truncated Wigner do occur at one loop, but only from coupled fields, or high momentum modes
when Truncated Wigner is implemented on a lattice. Otherwise, corrections are two-loops and higher order. 

All this is consistent with the way Truncated Wigner methods have been used so far in
particle and condensed matter physics. 
For example: in the theory of
quantum droplet mixtures, one-loop corrections for a Bose-Einstein condensate field
(called LHY corrections) are added to the field equation for the condensate \cite{Petrov2015}. 
However, the field equation is usually restricted to the so-called balanced case,
where the relative number of particles in each component is fixed. The loop corrections
come from the fluctuations orthogonal to the balanced combination. Therefore, this is an
example where the loop corrections come from integrating out a secondary field, and the Truncated
Wigner Approximation can be consistently applied to the remaining field \cite{Moss:2025xxq}.

Besides losing the one loop term, we shall also see that the two-loop corrections to Truncated Wigner
involve fewer Feynman diagrams than the full two-loop correction to the usual effective action. However, the corrections are
non-local, and require solving for the two-point function. This will be difficult to implement numerically,
but should be possible produce estimates, or maybe solve exactly in one dimension.

The consistency of Truncated Wigner approaches
to false vacuum decay has also been examined by comparing to $n$ particle irreducible methods in
\cite{Batini:2023zpi}. 
An alternative to Truncated Wigner for real time evolution and false vacuum decay can 
be found in \cite{Ai:2019fri}.
An extensive analytic examination of extensions to Truncated Wigner, including high momentum
modes as an environment and evolving a Fokker-Planck equation, can be found in \cite{Calzetta:2001pp}.

The next section contains an preview of the main results in the context of scalar field theory.
Section III contains the technical details of the effective action expansion, including corrections
from high frequency modes and from two-loop Feynman diagrams. We use units in which
$c=1$ and $\hbar=1$.

\section{Overview}

We leave the details for later and start with an overview of the results in the case of 
a Klein-Gordon scalar field $\phi$
with action
\begin{equation}
S[\phi]=\int d^3xdt\left(\frac12\dot\phi^2-\frac12(\nabla\phi)^2-V(\phi)\right).
\end{equation}
The strategy will be to introduce coherent states which depend on the field and its conjugate momentum.
An in-in effective action formalism is then used \cite{Chou:1984es,Calzetta:1986ey}, generated by
a path integral with two fields $\phi_1$ and $\phi_2$ with combined classical action 
$S[\phi_1,\phi_2]=S[\phi_1]-S[\phi_2]$. This path integral starts from an initial coherent state.
The effective field equations to be solved are
\begin{equation}
\frac{\delta\Gamma}{\delta \phi_1}=\frac{\delta\Gamma}{\delta \phi_2}=0,\qquad \phi=\phi_1=\phi_2,
\qquad \phi(x,t_0)=\phi_0(x),\ \dot\phi(x,t_0)=\pi_0(x).\label{eom2}
\end{equation}
In practice, we require evolution of the field operator expectation values $\langle\hat\phi\rangle$
from an initial density matrix $\rho$. Using coherent states, the density matrix
can be represented by a Wigner function $W(\phi_0,\pi_0)$. The results for the expectation values of
a field operator $\hat\phi(x,t)$ becomes
\begin{equation}
\langle\hat\phi(x,t)\rangle=\langle\phi(x,t)\rangle_W,
\end{equation}
where $\phi(x,t)$ is the solution to the effective field equations, and the expression
$\langle\rangle_W$ denotes a stochastic average over an ensemble
of paths with initial conditions that have a distribution $W(\phi_0,\pi_0)$.
(We will only consider gaussian initial states and then the Wigner function is positive definite.)
The correlation functions, on the other hand, have a correction
\begin{equation}
\langle\hat\phi(x,t)\hat\phi(x',t)\rangle=\langle\phi(x,t)\phi(x',t)\rangle_W+G^c(x,x',t)
\end{equation}
where $G^c$ vanishes at one loop order.

The above procedure is not new, in the sense that it has been used in a wide range
of applications. What we aim to do here is to put the formalism on a secure foundation,
and derive a loop expansion for the effective action $\Gamma$ including contributions from
the stochastic field.

\subsection{One loop}

The one loop approximation should give the first correction to the truncated Wigner formalism.
In the background field version, the fields in the path integral are displaced by a background value, 
$\phi_1\to\phi_1+\phi'_1$, $\phi_2\to\phi_2+\phi_2'$, and the path integrand expanded
to quadratic order in $\phi_1'$ and $\phi_2'$.
Propagator functions $G_{ij}$ are defined by a differential equation 
\begin{equation}
S^i{}_j[\phi_1,\phi_2]
\begin{pmatrix}
G_{11}(x,x')&G_{12}(x,x')\\
G_{21}(x,x')&G_{22}(x,x')\\
\end{pmatrix}
=
\begin{pmatrix}
-i\delta(x-x')&0\\
0&-i\delta(x-x')\\
\end{pmatrix}
\end{equation}
where $S^i{}_j$ is an operator formed from the second functional derivatives of the action,
\begin{equation}
S^i{}_j[\phi_1,\phi_2]=\begin{pmatrix}
-\nabla^2+V''(\phi_1)&0\\
0&\nabla^2-V''(\phi_2)\\
\end{pmatrix}
\end{equation}
Integrating out $\phi_1'$ and $\phi_2'$ from the path integral is a standard procedure, and the result is
the one-loop effective action $\Gamma[\phi_1,\phi_2]$,
\begin{equation}
\Gamma[\phi_1,\phi_2]=S[\phi_1,\phi_2]+\frac{i}{2}{\rm Tr}\,{\rm ln}\,S^i{}_j[\phi_1,\phi_2],\label{Gamma}
\end{equation}
where ${\rm Tr}$ is the functional trace of an operator. (This can be defined by a sum over the eigenvalues
of the operator with some regularisation procedure.)

The effective field equations are obtained by taking the variation of the action with respect to $\phi_1$
and then setting $\phi=\phi_1=\phi_2$. Using standard methods for differentiating the ${\rm Tr}\ln$ gives
\footnote{This equation can also be obtained by taking the expectation value of the operator field equation.}
\begin{equation}
-\nabla^2\phi+V'(\phi)+\frac12F(x,x)_RV^{\prime\prime\prime}(\phi(x))=0\label{phieq}
\end{equation}
where $F(x,x')$ is the symmetric propagator $\frac12(G_{11}(x,x')+G_{11}(x',x))$, and the 
$R$ subscript denotes regularisation that is necessary because $F(x,x)$ is infinite.
The function $F(x,x')$ satisfies
\begin{equation}
\left(-\nabla^2+V''(\phi(x)\right)F(x,x')=0.\label{Feq}
\end{equation}
Initial conditions are to be drawn from a Wigner function. We can choose to do this on $\phi$ and $\dot\phi$, 
or on the propagator.
Eq. (\ref{phieq}) should then be solved as a stochastic system, then averaged over the ensemble.

For example, consider an initial condition which is a pure gaussian state centred around a minimum of the potential
at $\phi_{\rm min}$. If we put the stochastic terms into the green functions, then the initial propagator is
\begin{equation}
F({\bf x},0,{\bf x}',0)=\int \frac{d^3k}{(2\pi)^3}\phi_k\phi_k^* e^{i{\bf k}({\bf x}-{\bf x}')}
\end{equation}
where $\phi_k$ and $\pi_k$ are drawn from the Wigner function. For an initial free-vacuum state, 
the Wigner function has gaussian form with expectation values
\begin{equation}
\langle |\phi_k|^2\rangle_W=\frac{1}{2\omega_k},\qquad
\langle |\pi_k|^2\rangle_W=\frac{\omega_k}{2},\label{Wav}
\end{equation}
and $\omega_k^2=k^2+V''(\phi_{\rm min})$. These initial conditions and field equations
are a simplified version of ones used in non-equilibrium quantum field theory \cite{Berges:2004yj}.

However, we could equally well put the stochastic terms in the initial values
of $\phi$ and $\dot\phi$, then the boundary conditions on $\phi_1'$ and $\phi_2'$ imply that $F=0$.
Since this is a non-trivial result, we put a more rigorous discussion in the next section.
The conclusion is that {\it the naive Truncated Wigner approach with the classical evolution is valid
at one loop in the effective action expansion}.

\subsection{Where have the one-loop corrections gone?}

In order to uncover the fate of the missing one-loop corrections in heuristic fashion
we will focus on $\lambda\phi^4$ theory
and make a few approximations.
The stochastic field satisfies
\begin{equation}
-\nabla^2\phi+m^2\phi+\frac16\lambda\phi^3=0,
\end{equation}
with gaussian random initial conditions. Take $\phi=\phi_c+\phi'$, where $\phi'$ is small
and contains all the randomness, together with $\langle\phi'\rangle=0$. The expectation
value of the field equation gives
\begin{equation}
-\nabla^2\phi_c+m^2\phi_c+\frac16\lambda\phi_c^3+
\frac12\lambda\phi_c\langle\phi^{\prime 2}\rangle_W=0\label{meaneq}
\end{equation}
Now suppose we keep the $\lambda\phi_c^2$ in the original equation, but drop all other $\lambda$ terms,
\begin{equation}
(-\nabla^2+m^2+\textstyle{\frac12}\lambda\phi_c^2)\phi'=0
\end{equation}
For slowly varying $\phi_c$, the Fourier components $\phi_k'(t)$ have approximate solution
\begin{equation}
\phi_k'(t)=\phi_k'(0)\cos\Omega_kt+\pi_k'(0)\frac{\sin\Omega_kt}{\Omega_k}
\end{equation}
where $\Omega_k^2=k^2+\frac12\lambda\phi_c^2$.
For a gaussian initial state with width given by $\Omega_k$ (rather than $\omega_k$) we get
\begin{equation}
\langle \phi^{\prime 2}\rangle_W=\int \frac{d^3k}{(2\pi)^3}\frac{1}{k^2+m^2+\lambda\phi_c^2/2}
\end{equation}
After regularisation, substituting into (\ref{meaneq}) gives a term similar to a one loop correction. 
If we use instead the gaussian initial state with width given by $\omega_k$, there is a correction which is
analytic in $\lambda\phi_c^2$. The leading term of the correction can be absorbed into the
renormalisation scale introduced by the regularisation procedure, and the remaining terms
are smaller that terms we have removed already.

\section{Complex field representations}

The classical evolution for a relativistic field requires setting initial conditions on
the field and conjugate momentum. This leads naturally to a path integral
formulation on phase space. The best way to quantise phase space is through the use 
of coherent states, and the corresponding complex representations for the fields.
Complex fields also arise naturally in non-relativistic field theory, with
the Bose-Einstein condensate field being the prime example.

Consider free Klein-Gordon theory with external source $J$, for example.
The action
\begin{equation}
S_J[\phi]=\int d^3xdt\left(\frac12\dot\phi^2-\frac12(\nabla\phi)^2-\frac12m^2\phi^2
-J\phi\right).
\end{equation}
We introduce complex
field components $\alpha_k(t)$, defined from the field and its conjugate momentum
$\pi=\dot\phi$,
\begin{align}
\phi&=\int d\hat k\left(\alpha_k e^{i{\bf k}\cdot{\bf x}}+
\overline\alpha_k e^{-i{\bf k}\cdot{\bf x}}\right)\\
\pi&=\int d\hat k\left(-i\omega_k\alpha_k e^{i{\bf k}\cdot{\bf x}}+
i\omega_k\overline\alpha_k e^{-i{\bf k}\cdot{\bf x}}\right).
\end{align}
The measure $d\hat k=d^3k/(8\pi^3\times2\omega_k)$. 
The action becomes
\begin{equation}
S_J[\alpha]=\int dtd\hat k\left\{
\frac{\dot{\overline\alpha_k}\alpha_k}{2i}-\frac{\dot\alpha_k\overline\alpha_k}{2i}
-{\cal H}_J,
\right\}
\end{equation}
where the Hamiltonian ${\cal H}_J$ depends on $\omega_k=(k^2+m^2)^{1/2}$,
\begin{equation}
{\cal H}_J=\alpha_k\overline\alpha_k\omega_k-J_k\overline\alpha_k-\overline J_k\alpha_k.
\end{equation}
Adding in interactions puts extra terms $H^I$ into the Hamiltonian. 
For $\lambda\phi^4$ theory,
\begin{equation}
H^I=\frac\lambda{24}\int d\hat k_1\dots d\hat k_4(2\pi)^3\delta(k_1+\dots k_4)
\prod_{i=1}^4(\alpha_{k_i}+\overline\alpha_{-k_i}).\label{HiR}
\end{equation}
When the field is quantised, the coefficients $\alpha_k$ and $\overline\alpha_k$
become annihilation and creation operators $a_k$ and $a_k^\dagger$. Coherent states
$|\alpha_k\rangle$ are eigenstates of the annihilation operators 
$a_k|\alpha_k\rangle=\alpha_k|\alpha_k\rangle$.

The difference between a relativistic system and a non-relativistic system occurs
in the definition of $\omega_k$. In the non-relativistic case $\omega_k=k^2/2m$,
and the measure is $d\hat k=d^3k/(2\pi)^3$. 
For a Bose-Einstein condensate with chemical potential $\mu$ 
and local interaction strength $g$, the $\alpha_k$ are the Fourier components of the
condensate field, and
\begin{equation}
H^I=-\mu\int d\hat k\,\alpha_k\overline\alpha_k+ \frac{g}{2}
\int d\hat k_1\dots d\hat k_4(2\pi)^3\delta(k_1+\dots k_4)
\alpha_{k_1}\alpha_{k_2}\overline\alpha_{k_3}\overline\alpha_{k_4}.\label{HiN}
\end{equation}

\subsection{Path integral formulations}

Consider a system initially described
by a density matrix $\rho$, which we can characterise by the coherent state matrix elements 
$\langle\alpha|\rho|\beta\rangle$. We shall be interested in how field expectation values
evolve. Consider the two point correlation, for example
\begin{equation}
\langle a_k(t)a^\dagger{}_{k'}(t)\rangle={\rm tr}\left(e^{iHt}a_ka^\dagger{}_{k'}e^{-iHt}\rho\right)
\end{equation}
Correlation functions like this can be obtained from functional differentiation of a generating function
\begin{equation}
Z_\rho[J,K]={\rm tr}\left(T^*e^{i\int H_Kdt}Te^{-i\int H_Jdt}\rho\right),
\end{equation}
where $T$ denotes time ordering, $T^*$ anti-time ordering and the subscripts denote the 
different source terms inserted into the Hamiltonian.
Inserting a decomposition of unity, this becomes
\footnote{We use a measure $d\alpha d\overline\alpha=id\alpha\wedge d\overline\alpha/2\pi$.}
\begin{equation}
Z_\rho[J,K]=
\int d\alpha d\overline\alpha d\beta d\overline\beta
\langle \beta|e^{iH_Kt}e^{-iH_Jt}|\alpha\rangle
\langle\alpha|\rho|\beta\rangle
\end{equation}
In the Schwinger-Keldysh path integral formalism, which we will be using,
the first matrix element is split into two paths, so that one set of fields
evolves from the initial to the final time ($\alpha_k$),  and the other from the final to the initial time ($\beta_k$).
The generating function is a path integral
\begin{equation}
Z_\rho[J,K]=\int D\alpha\,D\overline \alpha\,D\beta\,D\overline \beta\,\langle \alpha(t_0)|\rho|\beta(t_0)\rangle
\,e^{iS_J[\alpha]-iS_K[\beta]},
\end{equation}
where $S_J[\alpha]$ is the action for the path $\alpha(t)$ 
with external source $J$.

In order to simplify the notation, the $k$ will be omitted, and quadratic expressions
implicitly integrated over $k$. We introduce Keldysh variables $z$ and $\eta$ such that 
$\alpha=z+\eta/2$ and $\beta=z-\eta/2$.
For the exponential term, after expanding in powers of $\eta$ and integration by parts
\begin{equation}
S_{J}[\alpha]-S_{K}[\beta]=\frac{i}{2}\overline \eta(t_0)z(t_0)-\frac{i}{2}\overline z(t_0)\eta(t_0)
+S+S_I+S_J\label{exponent}
\end{equation}
where,
\begin{align}
S&=\int_{t_0}^\infty  dt\,\left\{\overline\eta\left(i\dot z-\frac{\partial H}{\partial \overline z}\right)-
 \left(i\dot{\overline z}+\frac{\partial H}{\partial z}\right)\eta\right\}\label{S0}\\
S_I&=\int_{t_0}^\infty dt \sum_{i+j odd}\frac{1}{2^{i+j-1}}\frac{1}{i!}\frac{1}{j!}
\overline\eta^i\frac{\partial^{i+j}{H}}{\partial \overline z^i\partial z^j}\eta^j\label{S1}\\
S_J&=\int_{t_0}^\infty dt \left\{\overline J_z z+\overline z J_z+\overline J_\eta \eta+\overline \eta J_\eta\right\}
\end{align}
In these formulae, the integral of $L$ is the sum of $L(t_0+n\delta t)\delta t$
starting from $n=1$, i.e $\eta(t_0)$ does not appear.

The density matrix can be replaced by a Wigner function $W(\alpha)$, using the definition
\begin{equation}
W(z)=\frac14\int d\eta d\overline\eta
\langle z+\eta/2|\hat\rho|z-\eta/2\rangle\,
e^{-\overline\eta z/2+\overline z\eta/2}
\end{equation}
Putting this together with (\ref{exponent}) and dropping an overall constant gives
\begin{equation}
Z_\rho[J_z,J_\eta]=\int Dz\,D\overline z\,D'\eta\,D'\overline \eta\,W(z(t_0))
\,\exp i\left(\int_{t_0}^\infty  dt\,\overline\eta(i\dot z-\frac{\partial H}{\partial \overline z})-
\int_{t_0}^\infty  dt (i\dot{\overline z}+\frac{\partial H}{\partial z})\eta+S_I+S_J\right)
\label{Zfull}
\end{equation}
where $D'$ indices that the integral at $t_0$ is omitted. 

Next, fix the initial condition $z(t_0)$ and define a new generating function $Z_0$,
\begin{equation}
Z_0[J_z,J_\eta]=\int D'z\,D'\overline z\,D'\eta\,D'\overline \eta
\,\exp i\left(\int_{t_0}^\infty  dt\,\overline\eta(i\dot z-\frac{\partial H}{\partial \overline z})-
\int_{t_0}^\infty  dt (i\dot{\overline z}+\frac{\partial H}{\partial z})\eta+S_I+S_J\right)
\end{equation}
The expectation values with initial condition $z_0$ are generated by functional derivatives
of $Z_0$, e.g
\begin{equation}
z\equiv\langle z_0| \hat z|z_0\rangle=\frac{1}{i}\frac{1}{Z_0}\frac{\delta Z_0}{\delta \bar J_z}
\end{equation}
The effective action for this initial condition is defined by
\begin{equation}
\Gamma[z,\bar z,\eta,\bar\eta]=\frac{1}{i}\ln Z_0-\int dt\left\{\bar J_z z+\bar z J_{\bar z}
+\bar J_\eta \eta+\bar \eta J_{\bar \eta}\right\}.
\end{equation}
Applying standard theoretical methods implies that the expectation values with no external 
source satisfy the effective field equations
\begin{equation}
\frac{\delta\Gamma}{\delta z}=\frac{\delta\Gamma}{\delta \eta}=0,\qquad \eta=0,\qquad z(t_0)=z_0.\label{eom}
\end{equation}
What we really desire is the evolution of $\langle\hat z\rangle$ from an initial density matrix $\hat\rho$, as generated by
$Z_\rho[J_z,J_\eta]$ in Eq. (\ref{Zfull}). Putting back the integrals over $z_0$ gives the main result,
\begin{equation}
\langle \hat z\rangle=\langle z \rangle_W,\label{aver}
\end{equation}
where $z(t)$ is a path that satisfies the effective equations of motion (\ref{eom}).
The expression $\langle\rangle_W$ denotes a stochastic average over an ensemble
of paths with initial conditions that have a distribution $W(z_0)$, where $W$ is the 
Wigner function for the density matrix $\hat\rho$. (Note that this only works if
$W(z_0)$ is positive definite, which is why we only consider gaussian initial distributions.)
The Truncated Wigner Approximation is obtained by setting $\Gamma=S$,
and the effective equations of motion become
\begin{equation}
i\dot z=\frac{\partial H}{\partial \overline z}
\end{equation}
The integral over $z(t_0)$ produces an ensemble of classical paths weighted by
the Wigner function. Higher loop corrections are considered later on.

\subsection{Correlators}\label{cors}

Equations (\ref{eom}) and (\ref{aver}) describe the evolution of the field expectation value in terms of a stochastic
system. We move on now to field correlators. Consider the
symmetric function with initial condition $z_0$ 
(also called the anti-commutator, Hadamard or statistical two-point function),
\begin{equation}
g_{kk'}(t,t')=\textstyle{\frac12}\langle z_0|\{\hat z_k(t),\hat {\overline z}_{k'}(t')\}|z_0\rangle\label{SymF}
\end{equation}
The symmetric function with initial density matrix $\rho$ is obtained by integrating over $z_0$,
\begin{equation}
G_{kk'}(t,t')=\langle g_{kk'}(t,t')\rangle_W
\end{equation}
We can rewrite this as
\begin{equation}
G_{kk'}(t,t')=\langle g^c_{kk'}(t,t')\rangle_W+\langle z_k(t)\overline z_{k'}(t')\rangle_W\label{corf}
\end{equation}
where $g^c$ is the connected symmetric function $g_{kk'}-z_kz_{k'}$  with initial condition $z_0$. The last term
in Eq. (\ref{corf}) can be obtained easily by evolving the field equations from stochastic initial
conditions and then averaging. The accuracy of this
procedure for obtaining the correlators depends on the extent to which the $g^c$ term can be neglected. 
We shall see later that it vanishes when using the Truncated Wigner Approximation, but contains loop
corrections beyond Truncated Wigner.

\subsection{Loop expansions in complex phase space}

In order to set this up, we will use DeWitt condensed notation \cite{DeWitt:1964mxt,Parker:2009uva},
where $\phi^I$ represents $(z_k(t),\bar z_k(t),\eta_k(t),\bar\eta_k(t))$ and
$\phi_I=\bar\phi^I$. A sum over $I$ implies a summation 
over all the indices, or integrals where these are non-discrete. The action of the theory can then be expanded
in a functional Taylor series about a background $\phi^I$,
\[
S[\phi^I+\sigma^I]=\sum_{n}\frac{1}{n\!}S_{I_1\dots i_n}[\phi^I]\sigma^{I_1}\dots\sigma^{I_n},
\]
where subscripts denote functional derivatives. Lines in the Feynman diagrams represent $iG^I{}_J$,
which is defined by connected expectation values of operators $\hat\sigma^I$ in a free theory with a 
background classical field $\phi^I$,
\[
iG^I{}_J=\langle z_0|\hat\sigma^I\hat\sigma_J|z_0\rangle_{\rm c}.
\]
The free theory is constructed by truncating the action after the $S_{IJ}$ term.
Functional derivatives of the generating function $Z_0$ give the relation between
$S_{IJ}$ and $G^{IJ}$,
\[
S^I{}_JG^J{}_K=\delta^I{}_K
\]
The use of Keldysh variables, in place of the original time-ordered $\alpha$ and $\beta$,
has implications for the causal structure of the propagators. If we decompose into the $z$
and $\eta$ sectors,
\[
G^I{}_J=
\begin{pmatrix}
F&G_R\\
G_A&0\\
\end{pmatrix},\label{decomp}
\]
where $G_R(t,t')$ represents expectation values which are only nonzero when
$t>t'$. This is therefore the retarded propagator. Similarly, we can show that $G_A$ is 
the advanced propagator, and $F$ is the symmetric function (\ref{SymF}) in the background field 
approximation.

The second order expansion term $S_{IJ}$, of the action (\ref{S0}), with a 
spatially constant background field becomes
\[
S^I{}_J=\begin{pmatrix}
0&\Delta\\
\Delta&0\\
\end{pmatrix}
,\qquad \Delta=
\begin{pmatrix}
i\partial _t-\omega_k-H^I_{z\bar z}&-H^I_{zz}\\
-H^I_{\bar z\bar z}&-i\partial_t-\omega_k-H^I_{z\bar z}\\
\end{pmatrix}
\]
where $z$ or $\bar z$ subscripts denote derivatives. We have fixed $z_0$ and $\bar z_0$.
The equation for $F$ is therefore
\begin{equation}
\Delta F=0,\qquad F(t,t_0)=F(t_0,t)=0.
\end{equation}
The solution is $F(t,t')=0$. On the other hand, we have
\begin{equation}
\Delta G_R=\delta(t-t'),\qquad G_R(t_0,t)=0,
\end{equation}
which has non-zero solutions. Similary for the advanced propagator $G_A$.

Earlier, in section \ref{cors}, we saw that the stochastic field evolution produced correlation functions that differed
from the quantum corellators by the connected function $g^c$. This function
is the $z\overline z$ component of the full connected green function. At leading order in the 
background field approximation, $g^c=F$, which vanishes. Therefore the stochastic approach
correctly reproduces the correlation functions for Truncated Wigner at one loop order.

\subsection{Constant backgrounds}

As an aside from the main developments in initial conditions, it is instructive to find the
propagator for constant backgrounds in the complex field representation, and to
relate this to the vacuum energy.
This is important because numerical methods used for Truncated Wigner
necessarily have a momentum cutoff, for example when using a spatial
lattice. The high momentum modes beyond the cutoff contribute to the effective action as a
correction term.
 
In frequency space, the second order expansion term of 
the action, $S_{IJ}$, with 
spatially constant background field becomes
\[
S^I{}_J=\begin{pmatrix}
0&\Delta\\
\Delta&0\\
\end{pmatrix}
,\qquad \Delta=
\begin{pmatrix}
\omega+\omega_k-H^I_{z\bar z}&-H^I_{zz}\\
-H^I_{\bar z\bar z}&-\omega-\omega_k-H^I_{z\bar z}\\
\end{pmatrix}
\]
where $z$ or $\bar z$ subscripts denote derivatives.
Note that
\begin{equation}
\det\Delta = -\omega^2+(\omega_k+H^I_{z\bar z})^2-
H^I_{zz}H^I_{\bar z\bar z}\equiv -\omega^2+\Omega^2_k,\label{OmegaDef}
\end{equation}
where we have defined a new frequency $\Omega_k$. Propagators can be obtained from the inversion of $S^I{}_J$. 
For constant background fields, the components of the propagator in Fourier space are explicitly
\begin{align*}
G_R&=\left[(\omega+i\epsilon)^2-\Omega_k^2\right]^{-1}\Delta'\\
G_A&=\left[(\omega-i\epsilon)^2-\Omega_k^2\right]^{-1}\Delta'
\end{align*}
where the $i\epsilon$ prescription is used to enforce the correct causality relations. The value
of $F$ depends on whether we put the initial conditions into the background field or the
fluctuations. Also,
\[
\Delta'=
\begin{pmatrix}
\omega-\omega_k-H^I_{z\bar z}&H^I_{zz}\\
H^I_{\bar z\bar z}&-\omega-\omega_k-H^I_{z\bar z}\\
\end{pmatrix}
\]
Note that the dispersion relation is $\omega=\Omega_k$, so that the vacuum energy is
\begin{equation}
\frac12\int \frac{d^3k}{(2\pi)^3}\,\Omega_k,
\end{equation}
independent of the value of $F$.

In the relativistic Klein-Gordon theory case, with potential $V$ and background field $\phi$, we have
\begin{equation}
H^I_{zz}=H^I_{\bar z\bar z}=H^I_{z\bar z}=\frac{1}{2\omega_k}(V''-m^2),
\end{equation}
where $\omega_k^2=k^2+m^2$. Hence, by Eq.(\ref{OmegaDef}),
\begin{equation}
\Omega_k^2=k^2+V''.
\end{equation}
Modes with momenta that lie between the Truncated Wigner cutoff 
$k_c$ and a regulator $\Lambda$ contribute a vacuum energy $E_{\rm extra}$,
\begin{equation}
E_{\rm extra}=\frac12\int_{k_c}^\Lambda\frac{d^3k}{(2\pi)^3}\left(k^2+V''\right)^{1/2}.
\end{equation}
In the case $k_c^2\gg V''$,
\begin{equation}
E_{\rm extra}=\frac{1}{16\pi^2}\left\{
[k^4]_{k_c}^\Lambda+V''[k^2]_{k_c}^\Lambda+\frac12(V'')^2\ln\frac{k_c}{\Lambda}
\right\}
\end{equation}
For $\lambda\phi^4$ theory, these terms can be absorbed into a renormalisation of the mass
and coupling in the original Hamiltonian, and then dropped when running the 
Truncated Wigner evolution for an initial vacuum state. However,
running the evolution with different cutoffs would require adjusting the mass. (Note that, in one
dimension, the correction is $\propto V''\ln(k_c/\Lambda)$. This effect has been
discussed previously as a cutoff dependent mass renormalisation in \cite{Braden:2022odm}.)

In the non-relativistic case, with (\ref{HiN}), constant background density $n$ and chemical potential $\mu=gn$,
\begin{equation}
H^I_{zz}=H^I_{\bar z\bar z}=H^I_{z\bar z}=gn.
\end{equation}
Hence, by Eq.(\ref{OmegaDef}),
\begin{equation}
\Omega_k^2=\omega_k(\omega_k+2gn),
\end{equation}
where $\omega_k=k^2/2m$. This gives a vacuum energy contribution to the Truncated Wigner evolution
of a Bose-Einstein condensate from modes beyond the cutoff,
\begin{equation}
E_{\rm extra}=\frac{1}{4\pi^2}\left\{
\frac1{10m}[k^5]_{k_c}^\Lambda+\frac13 gn[k^3]_{k_c}^\Lambda-mg^2n^2[k]_{k_c}^\Lambda
\right\}.
\end{equation}
These terms lead to cutoff dependence in the coupling used for Truncated Wigner. 
For an initial thermal state they would be temperature
dependent. In the case of a Bose-Einstein
condensate, there is a natural physical choice of cutoff such that that low momentum modes make 
up the condensate, and the high momentum modes make up an enveloping cloud.
This makes the cutoff temperature dependent.

\subsection{Beyond one loop}

The perturbative expansion of the effective action in powers of $\hbar$ is formally no different from
conventional quantum field theory (see, for example \cite{Parker:2009uva}),
\[
\Gamma=S+\frac{i}{2}\hbar\,{\rm tr}\,{\rm ln}\,S^I{}_J+i\sum_{L=2}^\infty\hbar^L(\hbox{1PI vacuum diagrams with L loops})
\]
The second term is real, despite appearances. (The simplest way to verify this is to perform a contour rotation
of the time integral $t\to i\tau$. The trace includes an integral over time, which is removes the factor of $i$, 
and also turns $S^I{}_J$ into a positive definite operator.)
\begin{center}
\begin{figure}[ht]
\begin{center}
\scalebox{0.15}{\includegraphics{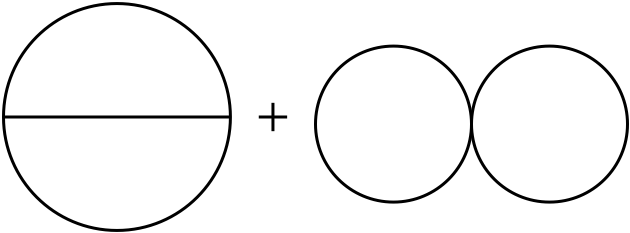}}
\end{center}
\caption{Two-loop contributions to the effective action. Lines represent free propagators
with a background field. Vertices represent derivatives of the action.}
\label{fig:twoloop}
\end{figure}
\end{center}

\begin{center}
\begin{figure}[ht]
\begin{center}
\scalebox{0.15}{\includegraphics{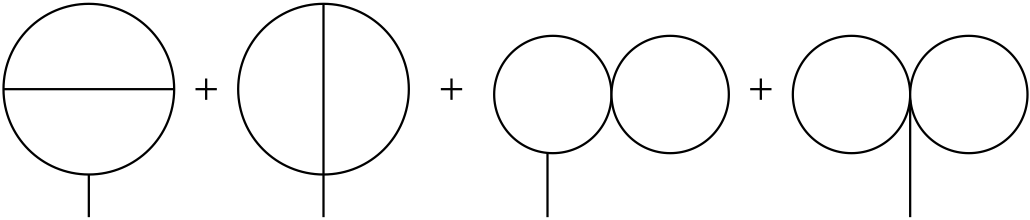}}
\end{center}
\caption{Two-loop contributions to the effective field equations.}
\label{fig:perturbative}
\end{figure}
\end{center}

The 1PI diagrams are 1 particle irreducible diagrams, meaning they cannot be decomposed into disconnected diagrams
by removing one line. The ends of all the lines must end on vertices, represented by the functional
derivatives  $iS_{I_1\dots I_n}$, $n>2$. Figure \ref{fig:twoloop} shows the two loop diagrams. Each diagram is 
accompanied by a  numerical symmetry factor.
The contribution of these two loop terms $\Gamma^{(2)}$ to the effective action is
\begin{equation}
\Gamma^{(2)}=\left\{\frac{1}{12}S_{IJK}S_{LMN}G^{IL}G^{JM}G^{KN}-\frac18S_{IJKL}G^{IJ}G^{KL}
\right\}
\end{equation}
This contributes to the effective field equation through differentiation with respect to $\phi^I$.
Vertices and propagators both depend on the field, and we have the identity
\begin{equation}
G^{IJ}{}_K=-G^{IL}S_{LMK}G^{MJ}.
\end{equation}
The effective field equation can be represented by tadpole diagrams, as in Figure 2.

The propagator decomposition (\ref{decomp}), with $F=0$, leads to a considerable simplification. The
decomposed propagator can be represented by placing arrows on the lines, going from the $z$ to the
$\eta$ field components. The vertices, from differentiating the action in Eqs. (\ref{S0}) and (\ref{S1}),
can be classified by the number of $z$ and $\eta$ legs,
\begin{equation}
(\hbox{no. of }z,\hbox{no. of }\eta)=(2,1),(3,1),(0,3),(1,3),\dots
\end{equation}
This leaves just one integral for each of the two-loop diagrams. Furthermore, closed loops
would represent $G_R(x,x)$, which vanishes. The tadpole diagrams that remain after decomposition are
shown in figure 3.

\begin{figure}[ht]
\begin{center}
\scalebox{0.2}{\includegraphics{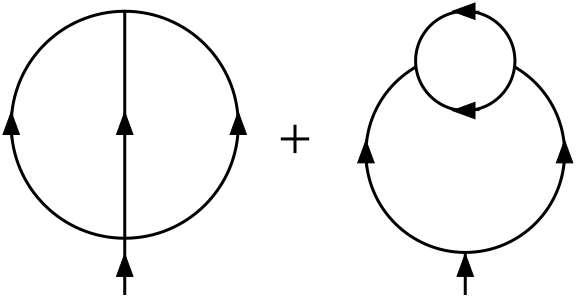}}
\caption{Two-loop contribution to the effective field equations after decomposing 
the propagators, with arrows indicating $z\to\eta$.}
\end{center}
\label{fig:perturbative}
\end{figure}

Consider $\lambda\phi^4$ theory with a real field representation. The
two-loop corrected form of the field equation is
\begin{align}
-\nabla^2\phi+m^2\phi&+\frac16\lambda\phi^3+
\frac16\lambda^2\int d^4x'G_R(x,x')^3\phi(x')\nonumber\\
&+\frac14\lambda^3\phi\int d^4x'd^4x''\,G_R(x,x')G_R(x,x'')G_R(x',x'')^2\phi(x')\phi(x'')=0,\label{2loop}
\end{align}
where the retarded propagator satisfies
\begin{equation}
(-\nabla^2+m^2+\textstyle{\frac12}\lambda\phi^2)G_R(x,x')=(2\pi)^4\delta(x-x').\label{GReq}
\end{equation}
The spacetime point $x=({\bf x},t)$, and time integrations range from $t_0$ to $t$, where $t_0$ is the initial time. 
Initial conditions for $\phi$ and $\dot\phi$ are drawn from a gaussian Wigner function distribution about some
initial value of $\phi$, with frequency $\omega_k$.

\begin{figure}[ht]
\begin{center}
\scalebox{0.2}{\includegraphics{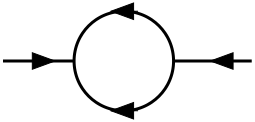}}
\caption{One-loop contribution to the symmetric function $F$ after decomposing 
the propagators, with arrows indicating $z\to\eta$.}
\end{center}
\label{fig:perturbative}
\end{figure}

The last correction term in Eq. (\ref{2loop}) contains the coincidence limit of
the one-loop correction $F^{(1)}$ to the  symmetric function,
given by Feynman diagram in Figure 4. The diagram corresponds to the integral 
\begin{equation}
F^{(1)}(x,y)=\frac12\lambda^2\int d^4x'd^4x''G_R(x,x')G_R(y,x'')G_R(x',x'')^2\phi(x')\phi(x'').\label{F1}
\end{equation}
In an earlier section, Sect. \ref{cors}, we discussed the relationship between the stochastic
correlator and the quantum correlation function. The difference was the connected symmetric function $g^c$,
which we now have to one-loop order in (\ref{F1}).

So far in this discussion, the corrections to the Truncated Wigner evolution have ignored the existence of a 
momentum cutoff in the numerical evolution. Including the cutoff gives a wide range of
extra terms that depend on $k_c$. In a renormalisable theory, we would expect to be
able to absorb the leading terms into the couplings of the theory.

The use of $\lambda\phi^4$ potentials is very limiting, especially if the aim is to study false vacuum
decay. A much better choice is to take a combination of $\cos\phi$ and $\cos 2\phi$ potentials with
two minima, one for a false vacuum and one for a true vacuum 
\cite{Braden:2017add,Braden:2018tky,Braden:2019vsw}. This type of potential has the advantage
of being renormalisable at one loop, which means contributions from modes above the cutoff
scale $k_c$ can be absorbed into the potential. The corrections follow the same pattern as above,
with vertices representing the derivatives of the potential. However, the $\lambda\phi^4$ potential
allows us to make some estimates of the loop terms in the next section.

\subsection{Secular growth}

Previous discussions of quantum corrections to the Truncated Wigner formalism in
optical lattices have
found that they grow quadratically with time \cite{PhysRevA.68.053604,Polkovnikov_2010}. 
We can examine if this is the case in the current formalism for $\lambda\phi^4$ quantum field theory by
isolating a single $k$ mode and putting the $\lambda=0$ solutions to the field and propagator
equations into the correction term. We also restrict the integrals to one
dimension. 

\begin{figure}[ht]
\begin{center}
\scalebox{0.3}{\includegraphics{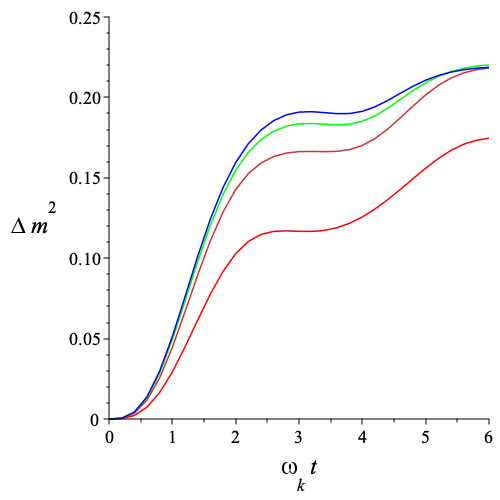}}
\caption{The integral for the first two-loop correction to Truncated Wigner
plotted as a function of time using the most basic approximations for the 
propagator and a field with momentum $k$.
The curves from red to blue show the results for $k=m,m/2,m/3,m/4$.
}\label{fig:int}
\end{center}
\label{fig:perturbative}
\end{figure}

Consider a harmonic field component initialised with amplitude $\phi_k$,
\begin{equation}
\phi_k(t)=\phi_k\sin\omega_k t.
\end{equation}
The $\lambda=0$ retarded propagator has Fourier components given by
\begin{equation}
G_R(t,t')=\frac{\sin\omega_k(t-t')}{\omega_k},\qquad t>t'.
\end{equation}
The first correction term to Eq. (\ref{2loop}) (with $t_0=0$) becomes 
$\frac16\lambda^2\phi_k\Delta$, where the integral $\Delta$ is
\begin{equation}
\Delta=\int_0^tdt'\int_{\sum k_i=k}\frac{d k_1}{2\pi}\frac{d k_2}{2\pi}\frac{d k_3}{2\pi}
\,\prod_i\left[\frac{\sin\omega_i(t-t')}{\omega_i}\right]\,\sin\omega_k t'
\end{equation}
and $\omega_i=(k_i^2+m^2)^{1/2}$. The time integral can be done first,
\begin{equation}
\Delta=\int_{\sum k_i=k}
d\hat k_1 d\hat k_2 d\hat k_3
\,f(\omega_1,\omega_2,\omega_3,\omega_k,t),
\end{equation}
where
\begin{equation}
f=\sum_{\tilde\omega\in S}\sigma_{\tilde\omega}
\frac{\sin\omega_kt-\sin\tilde\omega t}{\omega_k-\tilde\omega}.
\end{equation}
The set $S=\{\pm\omega_1\pm\omega_2\pm\omega_3\}$, and $\sigma_{\tilde\omega}=1$
if the number of minus signs is even, $-1$ otherwise.
There are apparent secular terms from the
integral near the resonant hypersurfaces $\tilde\omega=\omega_k$, where
$f=t+O(\tilde\omega-\omega_k)$.
This would cause the two-loop correction to grow linearly with time. However, after the
integrals over $k$ are completed, the growth becomes quadratic, and quickly saturates. This is shown
numerically for the one dimensional case in figure \ref{fig:int}.

\begin{figure}[ht]
\begin{center}
\scalebox{0.3}{\includegraphics{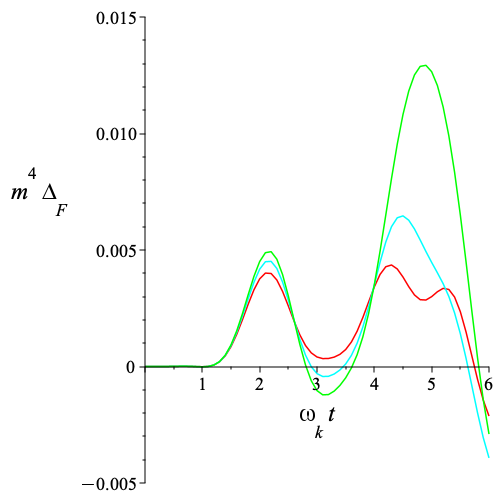}}
\caption{The integral for the loop correction to the equal time correlator using the most basic 
approximations for the propagator and a field with momentum $k=m$. The external momentum 
is $p=m$ (red), $p=m/2$ (cyan) and $p=0$ (green).
}\label{fig:F}
\end{center}
\label{fig:perturbative}
\end{figure}

The corrected symmetric function Eq. (\ref{F1}) determines the
difference between the stochastic and quantum field theory correlators.
Consider the correlation function at equal times, and Fourier transform 
the space dimension into a momentum $p$.
As above, the background will be a sine function with
momentum $k$,  so that 
$F^{(1)}(p,t,t)=\lambda^2\phi_k^2\Delta_F(p,k,t)$. Using the $\lambda=0$ green functions,
\begin{align}
&\Delta_F(p,k,t)=\int_0^t dt' \int_0^{t'}dt''\nonumber\\
&\int \frac{dk_1}{2\pi}\frac{dk_2}{2\pi}
\frac{\sin\omega_1(t''-t')}{\omega_1}\frac{\sin\omega_2(t''-t')}{\omega_2}
\frac{\sin\omega_k(t-t')}{\omega_k}\frac{\sin\omega_k(t-t'')}{\omega_k}
\sin\omega_kt'\sin\omega_kt'',
\end{align}
integrated over $k_1+k_2=k+p$.
This integral shows oscillatory growth with time, as shown in Fig. \ref{fig:F},
at a rate depending on the momentum. There is no sign of saturation, though this
may be a result of using a very crude approximation for the integrand. The results do show that
the stochastic correlators drift away from the true quantum field theory ones. This phenomenon
has been noticed before in Truncated Wigner simulations (e.g. \cite{Pirvu:2023plk}), where it has
been speculated that the fluctuations are equilibrating towards a Raleigh-Jeans thermal spectrum.
The loop corrections show the early stages of this drift in the correlators, and can be used to put
bounds on the error terms in the Truncated Wigner evolution.

\section{Conclusions}

The main conclusion is that it is consistent to initialise a quantum field with stochastic initial
conditions and integrate the field equations either classically, or including only the one-loop 
corrections from coupled fields. The same approximation that allows us to drop the one-loop
corrections of the stochastic field also leads to considerable simplification to the two-loop
and higher order terms.

The method applies equally well to relativistic and non-relativistic field theory. It is
especially useful for fields whose potential has multiple minima, where the calculation
of effective potentials is problematic. This includes the study of false vacuum decay,
where the approach has been widely used. It would be possible to go further in
these calculations and include two-loop effects, though this requires solving
propagator equations. The main application of our results may be to bound the size
of the correction terms.

Throughout the analysis, the choice of initial state has been restricted to a gaussian 
form, with a vacuum or false vacuum state in mind. Thermal, non-interacting, initial states 
can also be used, but thermal corrections from coupled fields or high momentum modes
then become increasingly important as the temperature is increased. These are the modes 
which can be dealt with most efficiently using a Langevin equation
\cite{Gleiser:1993ea,GardinerStochastic2002,GardinerStochastic2003}. 
The stochastic source in the Langevin equation takes into account the effects of leading 
order thermal corrections to the symmetric function $F$. Just like in the Truncated Wigner formalism, 
the stochastic equation includes only the classical propagators $G_R$ and $G_A$. The two approaches
are complementary, and can be combined at intermediate temperatures.

\acknowledgements

The author is grateful for discussions with Tom Billam and Alex Jenkins. This work
is supported by the Science and Technology
Facilities Council of the UK, grant ST/X000567/1.

\bibliography{paper.bib}

\end{document}